\documentclass[prb,showpaces,preprintnumbers,amsmath,amssymb,twocolumn,superscriptaddress]{revtex4-1}
\usepackage{graphicx}
\usepackage{dcolumn}
\usepackage{bm}
\usepackage{amssymb}
\usepackage{amsmath}
\usepackage{epstopdf}
\begin{document}
\title{Moderate interaction between $3d$- and $4f$-electrons and ferrimagnetism in Co-doped GdFeAsO}
\author{T. Shang}
\affiliation{Department of Physics and Center for Correlated Matter, Zhejiang University, Hangzhou, 310027, China}
\author{L. Yang}
\affiliation{Department of Physics and Center for Correlated Matter, Zhejiang University, Hangzhou, 310027, China}
\author{N. Cornell}
\affiliation{UTD-NanoTech Institute, The University of Texas at Dallas, Texas 75083$\textendash$0688, USA}
\author{F. Ronning}
\affiliation{Los Alamos National Laboratory, Los Alamos, New Mexico 87545, USA}
\author{Y. Chen}
\affiliation{Department of Physics and Center for Correlated Matter, Zhejiang University, Hangzhou, 310027, China}
\author{L. Jiao}
\affiliation{Department of Physics and Center for Correlated Matter, Zhejiang University, Hangzhou, 310027, China}
\author{J. Chen}
\affiliation{Department of Physics and Center for Correlated Matter, Zhejiang University, Hangzhou, 310027, China}
\author{A. Howard}
\affiliation{UTD-NanoTech Institute, The University of Texas at Dallas, Texas 75083$\textendash$0688, USA}
\author{J. D. Thompson}
\affiliation{Los Alamos National Laboratory, Los Alamos, New Mexico 87545, USA}
\author{A. Zakhidov}
\affiliation{UTD-NanoTech Institute, The University of Texas at Dallas, Texas 75083$\textendash$0688, USA}
\author{M. B. Salamon}
\affiliation{UTD-NanoTech Institute, The University of Texas at Dallas, Texas 75083$\textendash$0688, USA}
\author{H. Q. Yuan}
\email{hqyuan@zju.edu.cn}
\affiliation{Department of Physics and Center for Correlated Matter, Zhejiang University, Hangzhou, 310027, China}
\date{\today}

\begin{abstract}
We synthesized a series of GdFe$_{1-x}$Co$_x$AsO polycrystalline samples ($0 \leq x \leq 1$) by using a solid state reaction method and present a systematic study on the physical properties by means of electrical resistivity $\rho(T)$, magnetic susceptibility $\chi(T)$ and specific heat $C(T)$. The parent compound GdFeAsO undergoes a spin-density-wave (SDW) transition associated with Fe $3d$-electrons around 130 K, followed by an antiferromagnetic (AFM) transition of Gd at $T^\textup{Gd}_\textup{N} \approx$ 4 K. The SDW transition is quickly suppressed by Fe/Co substitution and superconductivity appears in a narrow doping range of $0.05 < x < 0.25$, showing a maximum $T_\textup{sc}$ $\approx$ 20 K around  $x = 0.1$. On the other hand, the $4f$-electrons of Gd are antiferromagnetically ordered over the entire doping concentration ($0 \leq x \leq 1$), while the Co $3d$-electrons exhibit a ferromagnetic (FM) transition above $x \approx 0.8$, with the Curie temperature ($T^\textup{Co}_\textup{C}$) reaching 75 K in GdCoAsO. These two magnetic species (Gd and Co) are coupled antiferromagnetically to give rise to ferrimagnetic behavior in magnetic susceptibility on the Co-rich side. For $0.7 \leq x < 1.0$, the system undergoes a possible magnetic reorientation below $T^\textup{Gd}_\textup{N}$.\\

\begin{description}
\item[PACS number(s)]
74.70.Xa,71.20.Eh,75.20.Hr
\end{description}
\end{abstract}

\maketitle

\section{\label{sec:level1}INTRODUCTION}
The discovery of superconductivity at $T_\textup{sc}$ = 26 K in LaFeAsO$_{1-x}$F$_x$ has stimulated intensive efforts on searching for new materials with higher superconducting transition temperature, $T_{sc}$, and revealing their unconventional nature.~\cite{1kamihara2008iron} Until now, a few series of iron-based superconductors (FeSCs) have been discovered, with a maximum $T_\textup{sc}$ raised to 56 K when La is substituted by other rare earth elements, e.g., Ce, Sm, Nd, Pr and Gd, in the $Ln$Fe$Pn$O family ($Ln$: lanthanoids, $Pn$: As or P).~\cite{2chen2008superconductivity, 3chen2008superconductivity2, 4Ren2008, 6wang2008thorium} The FeSCs show strong similarities to the copper oxides:~\cite{9armitage2010progress,9.1Yuri.Izyumov} (i) both possess a layered crystal structure and a relative high $T_\textup{sc}$; (ii) superconductivity seems to be closely tied up with magnetism. On the other hand, significant differences have been observed between them. For example, the parent compounds of FeSCs are bad metals, in contrast to the Mott-insulators in the cuprates.~\cite{9armitage2010progress,9.1Yuri.Izyumov} The superconducting order parameter of FeSCs is supposed to be S$_\pm$-type,~\cite{9.2mazin2008unconventional,Mazinreview} while the cuprates are d-wave superconductors.~\cite{Harlingen,9.3Tsuei} Moreover, the FeSCs show nearly isotropic upper critical fields at low temperature,~\cite{9.4yuan2009nearly} but the cuprates are rather anisotropic.~\cite{9.5worthington1987anisotropic} Electron doping into the conducting layers via elemental substitutions may induce superconductivity in iron pnictides, but it acts as a Cooper pair breaker in the cuprates.~\cite{9armitage2010progress,9.1Yuri.Izyumov} Thus, a detailed study of elemental substitutions on the Fe sites may help elucidate the mechanism of superconductivity in iron pnictides.

The $Ln$Fe$Pn$O materials consist of alternating $Ln$O and Fe$Pn$ blocks along the $c$-axis of a tetragonal ZrCuSiAs crystal structure.~\cite{9.1Yuri.Izyumov} The $Ln$O layers act as charge reservoirs while the Fe$Pn$ layers are conducting. Superconductivity can be induced by doping either electrons or holes into the $Ln$O layers, as well as doping electrons into the Fe$Pn$ layers.~\cite{9.1Yuri.Izyumov} GdFeAsO exhibits a structural/SDW transition around 130 K, followed by an AFM transition of Gd at 4.2 K.~\cite{6wang2008thorium} Partial substitution of Gd with Th suppresses the structural/SDW transition and then gives rise to superconductivity where $T_\textup{sc}$ reaches a maximum of 56 K at 20\% Th.~\cite{6wang2008thorium} Similarly, a narrow superconducting region was also observed in GdFe$_{1-x}$Ir$_x$AsO by Fe/Ir substitution~\cite{13cui2010superconductivity} and in GdFeAsO$_{1-\delta}$ by introducing oxygen deficiency.~\cite{12yang2008superconductivity}

In the past few years, considerable efforts have been devoted to the Co-doped $Ln$Fe$Pn$O compounds. However, these were mainly focused on superconductivity and magnetism in the Fe-rich materials. On the other hand,the Co-end compounds, namely $Ln$Co$Pn$O, demonstrate rich magnetic properties attributed to the interactions between the $3d$- and $4f$-electrons. For example, CeCoAsO and CeCoPO exhibit an enhanced Sommerfeld coefficient of 200 mJ/mol-K$^2$, and their Co-ions undergo a FM transition at $T^\textup{Co}_\textup{C}$ = 75 K.~\cite{23sarkar2010interplay, 24krellner2009interplay} LaCoAsO and LaCoPO are itinerant ferromagnets with a Curie temperature of $T^\textup{Co}_\textup{C} \simeq$ 50 K and 60 K, respectively.~\cite{22yanagi2008itinerant} Complex magnetic properties were observed in SmCoAsO and NdCoAsO, in which the Co-ions subsequently undergo a FM transition and an AFM transition upon cooling from room temperature,~\cite{25awana2010magnetic, 26mcguire2010magnetic} followed by an AFM transition of the $4f$-electrons at $T^\textup{Sm}_\textup{N}$ = 5 K and $T^\textup{Nd}_\textup{N}$ = 3.5 K.~\cite{25awana2010magnetic, 26mcguire2010magnetic} In GdCoAsO, both Co and Gd are magnetically ordered too,~\cite{27ohta2009magnetic} but there is no systematic study so far.

In order to study the interplay of $3d$- and $4f$-electrons and its associated consequences in iron pnictides, here we present a systematic study of the magnetism and superconductivity in GdFe$_{1-x}$Co$_x$AsO by means of measuring the electrical resistivity, magnetic susceptibility, isothermal magnetization and specific heat. A complete doping-temperature phase diagram is derived for $0\leq x \leq1$. It was found that the SDW transition of the Fe-ions is quickly suppressed by Fe/Co substitution and superconductivity occurs over a narrow doping range near the magnetic instability of Fe. On the other hand, the AFM order of Gd is rather robust against the Fe/Co substitution; it is hardly polarized even in the case that Co-ions are ferromagnetically ordered above $x \approx 0.8$. However, the coupling between the Gd- and Co-spices gives rise to ferrimagnetic behavior on the Co-side. We expect that these results, while revealing the emerging physics raised by the interplay of $3d$-$4f$ electrons, should be useful for further understanding the nature of superconductivity and magnetism in iron pnictides.

\section{EXPERIMENTAL DETAILS}
Polycrystalline samples of GdFe$_{1-x}$Co$_x$AsO ($0 \leq x \leq 1$) were synthesized in evacuated quartz ampoules by a two-step solid state reaction method. High-purity Gd (99.9$\%$) and As (99.999$\%$) chunks, also Co$_3$O$_4$ (99.97$\%$), Fe$_2$O$_3$ (99.99$\%$), Fe(99.9$\%$) and Co(99.9$\%$) powders were used as raw materials. The dehydrated Co$_3$O$_4$ and Fe$_2$O$_3$ were prepared by heating the powders at 1173 K for 10 h in atmosphere. First, Gd and As chunks were mixed and the pure phase GdAs was prepared by heating the mixture in a sealed evacuated quartz ampoule at 1073 K for 24 h, and then at 1323 K for 48 h. Then, GdAs, Co$_3$O$_4$, Fe$_2$O$_3$, Fe and Co powders were weighed according to the stoichiometric ratio, thoroughly ground, and pressed into pellets. The entire process was carried out in an argon-filled glove box. Finally, the pellets were placed in alumina crucibles and sealed into evacuated quartz ampoules, then slowly heated to 1323 K at a rate of 100 K/h and kept at this temperature for one week to obtain sintered pellets.

The structure of these polycrystalline samples was characterized by powder X-ray diffraction (XRD) at room temperature using a PANalytical X'Pert MRD diffractometer with Cu K$\alpha$ radiation and a graphite monochromator. Measurements of the DC magnetic susceptibility, electrical resistivity and specific heat were carried out in a Quantum Design Magnetic Property Measurement System (MPMS-5) and the Physical Property Measurement System (PPMS-9), respectively. Temperature dependence of the electrical resistivity was measured by a standard four-point method.

\section{RESULTS AND DISCUSSION}
\subsection{\label{sec:level2}Crystal Structure}

\begin{figure}[tbp]
     \begin{center}
     \includegraphics[width=3.35in,keepaspectratio]{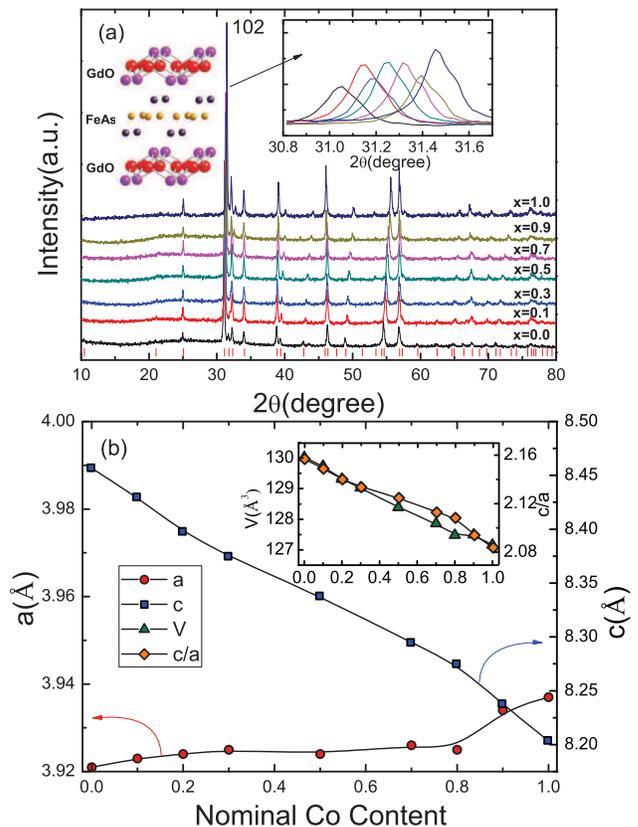}
     \end{center}
     \caption{(Color online) Powder X-ray diffraction patterns and the lattice parameters of GdFe$_{1-x}$Co$_x$AsO. (a) Room temperature XRD patterns of $x = 0.0, 0.1, 0.3, 0.5, 0.7, 0.9, 1.0$. The left inset shows the crystal structure and the right one enlarges the (\textbf{102}) peaks. (b) Lattice parameters as a function of nominal Co content. The inset plots the unit cell volume (left axis) and the $c/a$ ratio (right axis).}
     \label{fig1}
\end{figure}

Figure 1(a) plots seven representative XRD patterns of GdFe$_{1-x}$Co$_x$AsO ($0 \leq x \leq 1$) polycrystalline samples, the other samples showing similar patterns. The vertical bars on the bottom denote the calculated positions of Bragg diffraction of GdFeAsO. All the peaks can be well indexed based on the ZrCuSiAs tetragonal structure with the space group P4/nmm, as shown in the left inset of Fig.1(a). No obvious impurity phases can be detected, indicating the high quality of these samples. The shift of the (\textbf{102}) peak toward the right (larger $2\theta$ value) with increasing $x$, as shown in the right inset of Fig.1(a), reveals the contraction of crystal lattice with increasing Co concentration. The lattice parameters, refined by the least square fitting for GdFeAsO: $a = 3.921 \textup{\r{A}}$ $c = 8.4570 \textup{\r{A}}$ and GdCoAsO: $a = 4.937 \textup{\r{A}}$ $c = 8.204 \textup{\r{A}}$, are in good agreement with reported structural data.~\cite{6wang2008thorium, 27ohta2009magnetic} Figure 1(b) plots the lattice parameters as a function of nominal Co concentration. The $c$-axis shrinks significantly while the $a$-axis increases slightly with increasing Co concentration, which means that Fe/Co substitution increases the three dimensionality. This becomes more pronounced above $x = 0.8$, for which the Co-electrons are ferromagnetically ordered at low temperatures. The shrinkage of the $c$-axis reduces the distance between GdO and Fe(Co)As layers, thus, enhances the $3d$-$4f$ exchange interactions. The inset of Fig.1(b) shows the unit cell volume and the $c/a$ ratio versus Co concentration, both of which decrease with increasing $x$. The unit volume decreases by 2.20$\%$ at $x = 1$. The shift of the XRD patterns and also the corresponding variation of the lattice parameters suggest that Co atoms are successfully incorporated into the lattice. These results are consistent with other reports on the Co-doped LaFeAsO, CeFeAsO and NdFeAsO compounds.~\cite{16sefat2008superconductivity, 17tian, 18kim2010growth}

\subsection{Electrical Resistivity}
Figure 2 plots the temperature dependence of the normalized electrical resistivity of GdFe$_{1-x}$Co$_x$AsO ($0 \leq x \leq 1$) at zero-field. The resistive anomaly observed around $T_\textup{SDW} \approx$ 130 K in GdFeAsO characterizes the almost degenerate SDW transition of Fe $3d$-electrons and the tetragonal to orthorhombic structural phase transition.~\cite{6wang2008thorium,31dong2008competing,32zhao2008structural} Upon further decreasing temperature, a resistive kink appears around 3.6 K, as shown by the arrow in the left inset of Fig.2(a), below which the Gd $4f$-electrons become antiferromagnetically ordered.~\cite{6wang2008thorium} Upon partially substituting Fe with Co, the resistive anomaly at $T_\textup{SDW}$ is quickly suppressed and becomes imperceptible around $x = 0.07$. Meanwhile, superconductivity appears around $x$ = 0.07, and the transition temperature reaches a maximum of $T{^\textup{{onset}}_\textup{sc}} \approx 20$ K around $x = 0.1$ (see the right inset of Fig.2(a)). Note that, for $0.07 \leq x \leq 0.2$, no sign of an AFM transition can be identified for Gd $4f$-electrons in the electrical resistivity, being different from that of CeFeAsO$_{1-x}$F$_x$ in which a non-zero resistive transition shows up below $T_\textup{sc}$ attributed to the magnetic order of Ce.~\cite{tian} However, evidence for the AFM transition of Gd can be inferred from the magnetic susceptibility (see below). For $x = 0.25$, no superconductivity can be observed down to 3 K, and the AFM transition of Gd shows up around 4 K. In comparison with the Th-doped GdFeAsO,~\cite{6wang2008thorium} the most obvious difference, other than the lower $T_\textup{sc}$, is the semiconducting behavior observed in the under-doped GdFe$_{1-x}$Co$_x$AsO ($x < 0.1$). Similar behavior was also observed in CeFe$_{1-x}$Co$_x$AsO compounds,~\cite{17tian} which can be ascribed to either doping-induced disorders or Kondo-like impurities within the FeAs layers. Further studies are needed to clarify this issue.

\begin{figure}[tbp]
     \begin{center}
     \includegraphics[width=3.35in,keepaspectratio]{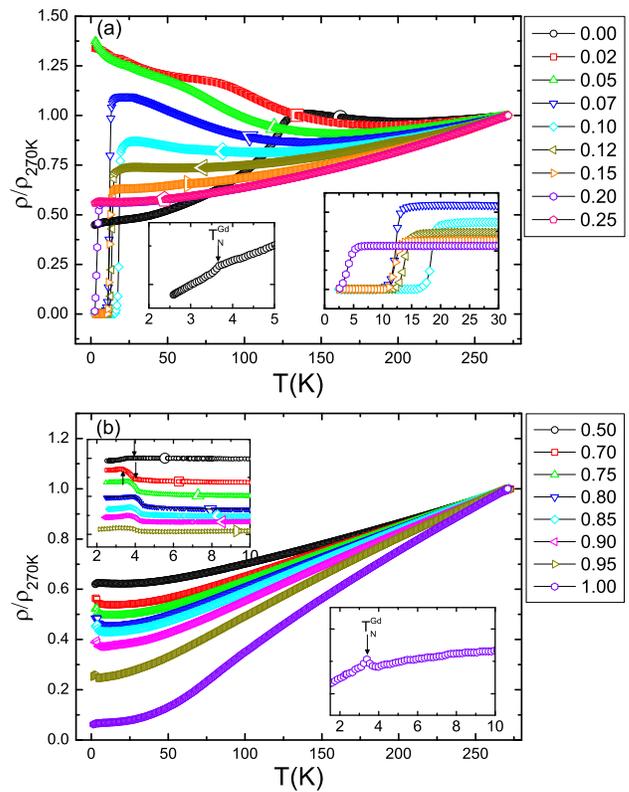}
     \end{center}
     \caption{(Color online) Temperature dependence of the normalized electrical resistivity of GdFe$_{1-x}$Co$_x$AsO. (a) $0.0 \leq x \leq 0.25$. (b) $0.5 \leq x \leq 1.0$.  The insets expand the low temperature regime.}
     \label{fig2}
\end{figure}

Figure 2(b) presents the normalized electrical resistivity for the heavily Co-doped samples ($0.5 \leq x \leq 1.0$), which decreases monotonically  with decreasing temperature, showing metallic behavior above 5 K. The electrical resistivity demonstrates a stronger temperature dependence for $x \geq 0.85$, which is likely attributed to the FM order of Co $3d$-electrons, even though such a transition is weak in the electrical resistivity. Again, the Gd $4f$-electrons undergo an AFM transition below 4 K. In particular, a step-like upturn transition appears at low temperature for $0.7 \leq x \leq 0.95$, which can be clearly seen from the upper inset of Fig.2(b). For example, the electrical resistivity for $x = 0.7$ increases abruptly at 3.9 K and then decreases below 3.2 K. Such a resistive upturn might be associated with a gap opening at the AFM transition temperature of Gd. The subsequent resistive decrease may arise from a magnetic reorientation, which is observed in the bulk properties too (see below). This step-like transition disappears in the stoichiometric GdCoAsO material, in which only a small kink associated with the AFM transition of Gd $4f$-electrons is observed around $T^\textup{Gd}_\textup{N}$ $\approx$ 3.5 K, as shown in the lower inset of Fig.2(b).

\begin{table}[tbp]
\centering
\caption{\label{tab:table1} The superconducting onset temperatures $T_\textup{sc}$(K) for $0.07 \leq x \leq 0.2$, determined from the 90$\%$ of the normal resistivity at $T_\textup{sc}$ ($T{^\rho_\textup{sc}}$) and the onset of magnetic susceptibility ($T{^\chi_\textup{sc}}$).\\}
\begin{ruledtabular}
\begin{tabular}{lcc}
\textrm{x(Co concentration)}&
\textrm{$T{^\rho_\textup{sc}}$ (K)}&
\textrm{$T{^\chi_\textup{sc}}$ (K)}\\
\colrule
0.07 & 14 & 13\\
0.10 & 20 & 19\\
0.12 & 15 & 15\\
0.15 & 13 & 11\\
0.20 & 6  & No\\
\end{tabular}
\end{ruledtabular}
\end{table}

\begin{figure}[tbp]
     \begin{center}
     \includegraphics[width=3.35in,keepaspectratio]{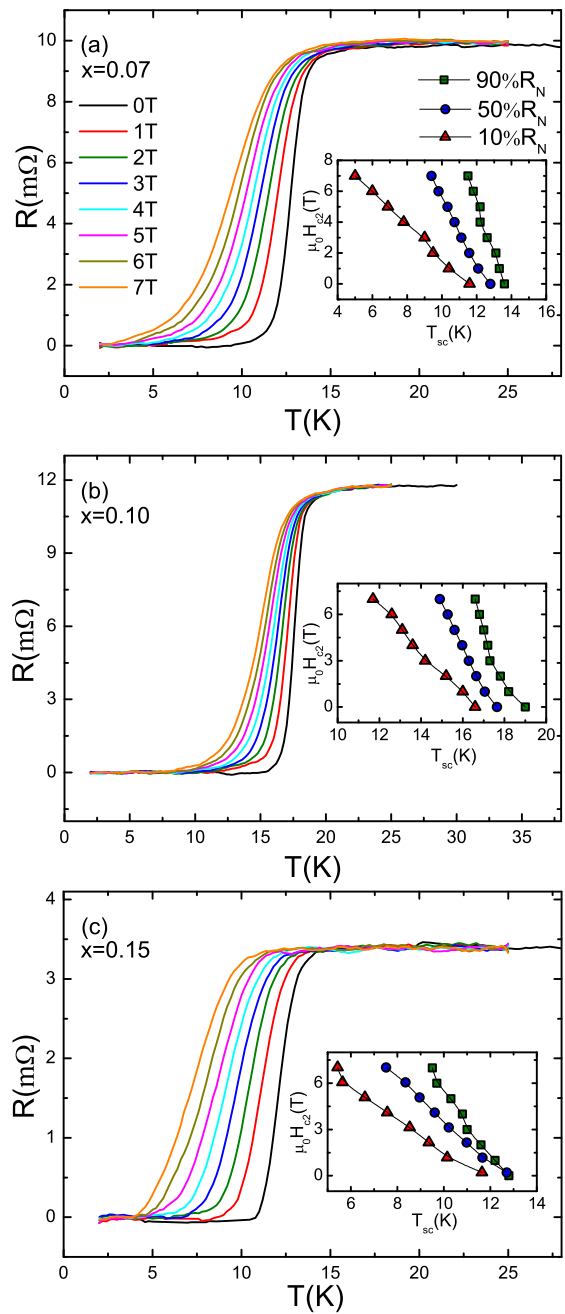}
     \end{center}
     \caption{(Color online)  Temperature dependence of the electrical resistance at various magnetic fields for $x$ = 0.07 (a), 0.10 (b), 0.15 (c). The insets show the upper critical fields $\mu_0 H_\textup{c2}(T_\textup{sc})$ obtained at the 90$\%$ (green square), 50$\%$ (blue circle), 10$\%$ (red triangle) of the normal resistance R$_\textup{N}$.}
     \label{fig3}
\end{figure}

Figure 3 presents the resistive superconducting transitions for $x$ = 0.07, 0.10 and 0.15 at various magnetic fields. In zero field, all the samples show a sharp transition with $T{^\textup{{mid}}_\textup{sc}}$ $\approx$ 12.8 K, 17.6 K and 12.2 K for $x$ = 0.07, 0.10 and 0.15, respectively. Application of a magnetic field shifts the superconducting transition to lower temperatures and significantly broadens the transition width. The latter is likely attributed to the vortex flow as previously discussed in other Fe-based superconductors.~\cite{jiao, aroszynski, Lee} The insets of Fig.3 plot the upper critical fields $\mu_0 H_\textup{c2}(T_\textup{sc})$, which were derived at various resistive drops of the superconducting transition. Following the WHH model: $\mu_0 H_{c2}(0)=-0.693T_\textup{sc}[dH_\textup{c2}/dT]$,~\cite{29werthamer1966temperature} one can estimate $\mu_0$$H_\textup{c2}$(0), which gives $\mu_0$$H_\textup{c2}$(0) = 19 T, 32 T  and 13 T for $x$ = 0.07, 0.1 and 0.15, respectively. The coherence length $\xi(0)$ can be calculated from the Ginzburg-Landau formula $\xi(0) \approx (\Phi_{0}/2 \pi H_{c2})^{1/2}$, where $\Phi_0$ is the quantum of magnetic flux. This yields $\xi(0) = 41 \textup{\r{A}}, 32 \textup{\r{A}}$ and 50 $\textup{\r{A}}$ for $x = 0.07, 0.10$ and 0.15, respectively.

\subsection{Magnetic Properties}
DC magnetization of GdFe$_{1-x}$Co$_x$AsO was measured as functions of temperature and magnetic field. Figure 4 shows the low temperature magnetic susceptibility $\chi(T)$ for the superconducting samples, measured in a field of $\mu_0 H$ = 2 mT. Evidence of bulk superconductivity was observed for $0.07 \leq x \leq 0.2$ samples. The superconducting volume fraction reaches over 50$\%$ for $x$ = 0.1, 0.12 and 0.15. Furthermore, a magnetic transition can be tracked around 4 K as denoted by the arrow in the inset, which marks the magnetic transition of Gd inside the superconducting state. The superconducting transition temperatures $T_{sc}$, derived from the electrical resistivity and magnetic susceptibility, are summarized in Table I.

\begin{figure}[tbp]
     \begin{center}
     \includegraphics[width=3.35in,keepaspectratio]{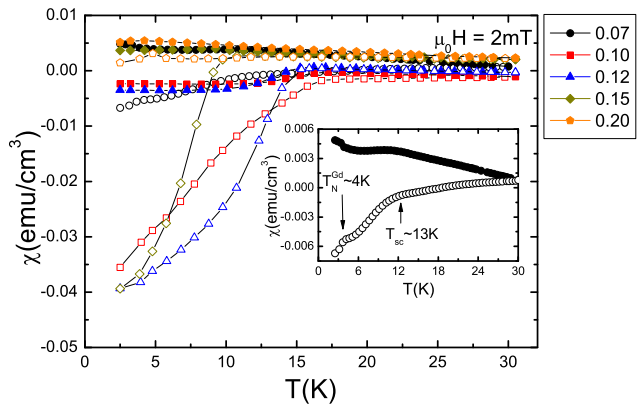}
     \end{center}
     \caption{(Color online) Low temperature magnetic susceptibility $\chi(T)$ for GdFe$_{1-x}$Co$_x$AsO, $0.07 \leq x \leq 0.20$. The open and filled symbols show the data measured at zero-field cooling (ZFC) and field cooling (FC), respectively. The inset shows the case for $x = 0.07$, where the AFM transition of Gd was observed below the superconducting transition at $T_\textup{sc}$ = 13 K.}
     \label{fig4}
\end{figure}
\begin{figure}[tbp]
     \begin{center}
     \includegraphics[width=3.35in,keepaspectratio]{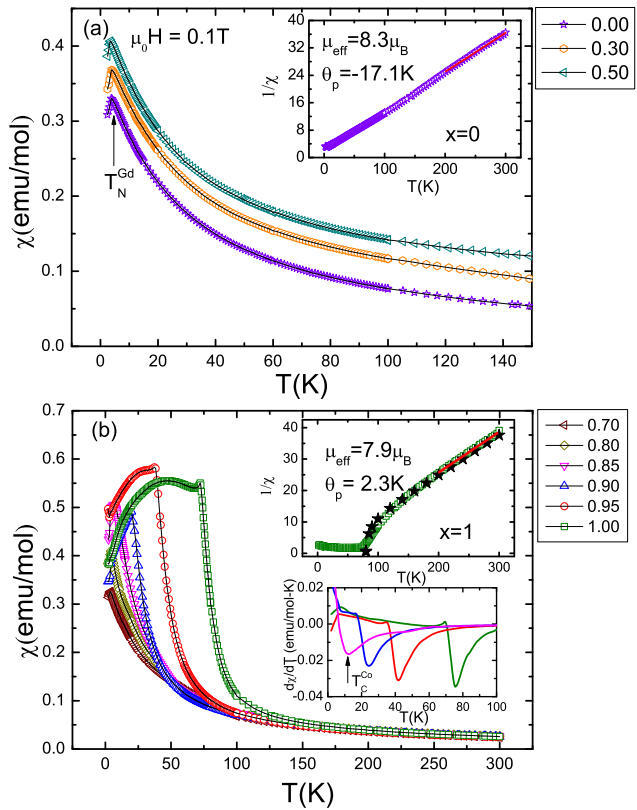}
     \end{center}
     \caption{(Color online) Temperature dependence of the DC magnetic susceptibility of GdFe$_{1-x}$Co$_x$AsO, $0 \leq x \leq 1$. No difference was observed for the ZFC- and FC-data. Note that the curves for $x = 0.3$ and $x = 0.5$ have been shifted by an offset of 0.05 emu/mol. The insets plot the inverse susceptibility as a function of temperature for $x = 0$ (a) and $x = 1$ (b). The red lines are fits to the Curie-Weiss law and the black stars are a fit to the mean field model.}
     \label{fig5}
\end{figure}
\begin{figure*}[tbp]
     \begin{center}
     \includegraphics[width=7in,keepaspectratio]{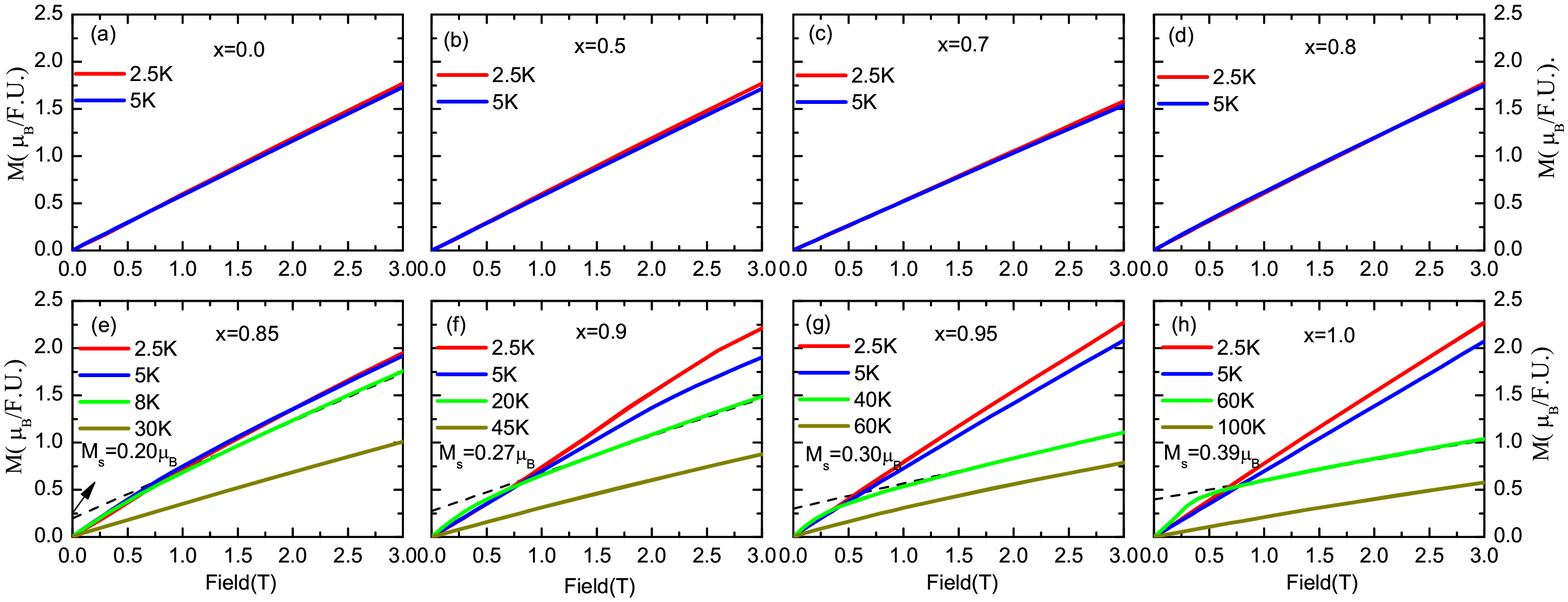}
     \end{center}
     \caption{(Color online) Field dependence of the magnetization M($H$) at various temperatures for GdFe$_{1-x}$Co$_x$AsO, $0 \leq x \leq 1$.}
     \label{fig6}
\end{figure*}

Magnetic susceptibility of the non-superconducting samples, measured in a field of $\mu_0 H$ = 0.1 T, is presented in Fig.5.
To make it clear, a value of 0.05 emu/mol is added as an offset in Fig.5(a). For $x \leq 0.7$, the magnetic susceptibility $\chi(T)$ shows a peak around 4 K, which corresponds to the AFM transition of Gd, and is nearly independent of the Co doping concentration. In the high temperature range ($T >$ 200 K), the magnetic susceptibility $\chi(T)$ can be well described by the Curie-Weiss law: $\chi(T) = \chi_0 + C/(T - \theta_p)$, with $\chi_0$ being a temperature-independent susceptibility, $C$ the Curie constant and $\theta_p$ the paramagnetic Curie temperature. In the inset of Fig.5(a), we plot 1/$\chi(T)$ versus $T$ for GdFeAsO; the red line is a fit to the Curie-Weiss law. The derived effective moment and paramagnetic Curie temperature are 8.3$\mu_B$ and -17.1 K, respectively. This effective moment is close to but somewhat lager than the free-ion moment of Gd, possibly due to the contributions of $5d$ electrons.\cite{moon} As the Fe/Co substitution increases beyond $x = 0.7$, the behavior changes dramatically. The susceptibility, as shown in Fig.5(b), rises sharply at a temperature that increases with $x$, reaching 75 K at $x = 1$.  The inverse susceptibility for $x = 1$ can be also fitted by the Curie-Weiss law at the temperature above 200 K (see inset of Fig.5(b)), with a slope consistent with a magnetic moment of 7.9 $\mu_B$, very close to the free-ion moment of Gd. However, the value of $\theta_p$ = 2.3 K, is substantially smaller than that of the $x = 0$ sample (Fig.5(a) inset). The downward curvature of the inverse susceptibility is characteristic of a ferrimagnet,~\cite{Sugiyama} in which two magnetic species are coupled antiferromagnetically. The strong dependence of the transition temperature $T^\textup{Co}_\textup{C}$ with Co concentration suggests that the FM coupling among Co-atoms, rather than strong AFM coupling to the Gd, is the primary driver of the transition.

We model the behavior within a mean field approach,~\cite{33Kittel} in which the magnetization of Co atoms is given by $M_\textup{Co}T=C_\textup{Co}(H+\lambda M_\textup{Co}-\mu M_\textup{Gd})$, and that of the Gd by $M_\textup{Gd}T=C_\textup{Gd}(H-\mu M_\textup{Co})$. Here $C_\textup{Co}$($C_\textup{Gd}$) is the Curie constant for Co(Gd), $\lambda$ is the FM Co-Co coupling constant and $\mu$ is the AFM Co-Gd coupling. We assume that $C_\textup{Gd}$ follows from the full atomic moment of Gd, 7.9$\mu_B$, and seek a solution in which $T^\textup{Co}_\textup{C} \approx \lambda C_\textup{Co}$  and $\theta_p \approx 2.3$ K.  The solid star symbols in the inset of Fig.5(b) show a fit to the data with $\mu / \lambda \approx 0.02$, confirming that the AFM coupling between the two species is relatively weak.

Despite the weak coupling, the large magnetic moment of Gd leads to an appreciable magnetization that opposes that of the Co, causing a strong decrease in the combined magnetization at low temperature. The data in Fig.6 confirm this picture, which plots the field dependence of magnetization at different temperatures with the magnetic field up to 2 T. Below $x = 0.85$, the magnetization is strictly linear in field and independent of Co concentration. For $x \geq 0.85$, in which the Co moments are ferromagnetically ordered, spontaneous magnetization was observed below the Curie temperature $T^\textup{Co}_\textup{C}$ and its size increases with increasing the Co concentration. For GdCoAsO, the magnetization reaches 0.39 $\mu_B$ with a maximum near 60 K. Similar results were also reported in LaCoAsO, NdCoAsO and SmCoAsO compounds with a saturated moment of 0.46$\mu_B$, 0.20$\mu_B$ and 0.18$\mu_B$, respectively.~\cite{22yanagi2008itinerant, 25awana2010magnetic, 26mcguire2010magnetic} The saturated moments of Co in these compounds are much smaller than 3$\mu_B$/Co$^{2+}$ as expected for the localized high spin Co$^{2+}$ ion, indicating that the ferromagnetism in the Co layer is itinerant in nature. The fit shown in Fig.5, however, suggests that the Co moment is closer to 0.5$\mu_B$, indicating that the spontaneous magnetization in Fig.6 is reduced by the negative magnetization induced in the Gd. At low temperatures, the spontaneous magnetization has disappeared and the magnetization is again linear in field. We expect that the Gd blocks order antiferromagnetically at low temperature and tend to align the magnetically-ordered Co blocks oppositely.  The type of behavior has been observed by means of neutron scattering in NdCoAsO.~\cite{26mcguire2010magnetic} In that case, the AFM interactions among the Nd atoms and between Nd and Co are sufficiently strong enough to reorient the FM Co layers into an antiparallel alignment.

\subsection{Heat Capacity}
In order to further characterize the magnetic order of Gd 4$f$-electrons in GdFe$_{1-x}$Co$_x$AsO, we also performed measurements of the low temperature specific heat $C(T)$ at zero-field. Figure 7 shows the magnetic susceptibility $\chi(T)$ and the specific heat $C(T)/T$ of several representative doping concentrations. For $x < 0.7$, observations of a $\lambda$-type anomaly in specific heat and a cusp in magnetic susceptibility unambiguously identify the AFM transition of Gd around $T_\textup{N}^\textup{Gd} \approx 4$ K. With increasing the Co content, a second anomaly develops in the heat capacity at lower temperature for $x \geq 0.7$, but disappears again at $x = 1$. Similar evidence can be also inferred from the magnetic susceptibility $\chi(T)$, which is most clearly seen in $x = 0.85$. As already described above, the second transition was also observed in the electrical resistivity (see Fig.2(b)). All these indicate that the system ($0.7 \leq x < 1.0$) may undergo a magnetic reorientation transition below $T_\textup{N}^\textup{Gd}$ attributed to the coupling between the Co $3d$- and Gd $4f$-electrons.

\begin{figure}[tbp]
     \begin{center}
     \includegraphics[width=3.35in,keepaspectratio]{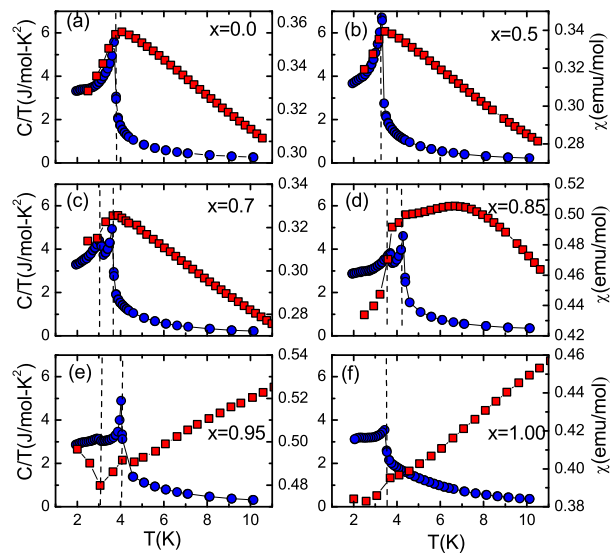}
     \end{center}
     \caption{(Color online) Temperature dependence of the specific heat $C(T)$ (blue circle, left axis) and magnetic susceptibility $\chi(T)$ (red square, right axis) of GdFe$_{1-x}$Co$_x$AsO, $0 \leq x \leq 1$.}
     \label{fig7}
\end{figure}
\subsection{Phase Diagram and Discussion}
\begin{figure}[tbp]
    \begin{center}
    \includegraphics[width=3.35in,keepaspectratio]{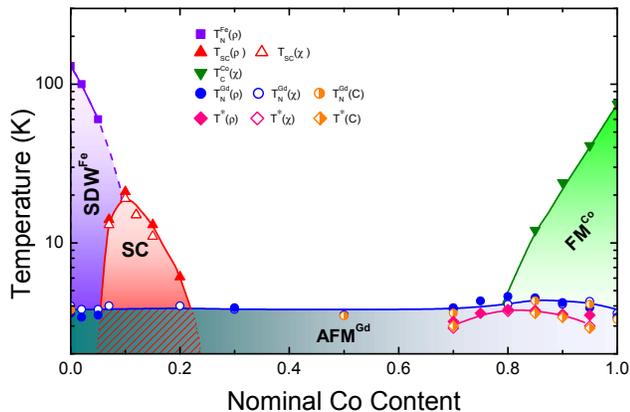}
    \end{center}
    \caption{(Color online) Magnetic and superconducting phase diagram of GdFe$_{1-x}$Co$_x$AsO as a function of nominal Co concentration. Various symbols denote the different ordering temperatures. The solid lines are guides to the eyes.}
    \label{fig8}
\end{figure}

Based on the above experimental data, we present a magnetic and superconducting phase diagram of GdFe$_{1-x}$Co$_x$AsO ($0 \leq x \leq 1$) as a function of nominal Co concentration, as plotted in Fig.8. The parent compound GdFeAsO is a bad metal, with the Fe $3d$- and Gd $4f$-electrons ordered antiferromagnetically below 130 K and 4 K, respectively. Upon Fe/Co substitution, the SDW transition of Fe $3d$-electrons vanishes around $x = 0.1$ (marked by SDW$^\textup{Fe}$) and superconductivity appears in a range $0.05 < x < 0.25$, with the highest $T_\textup{sc}$ $\approx$ 20 K around $x =0.1$ (marked by SC). This dome-like superconducting region is close to the SDW instability, indicating that superconductivity is related to magnetism in these compounds. At lower temperature, Gd is antiferromagnetically ordered with $T^\textup{Gd}_\textup{N}$ $\approx$ 4 K (marked by AFM$^\textup{Gd}$), being nearly independent of the Fe/Co substitution. The Gd AFM order coexists with the Fe SDW order for $x < 0.07$ and survives in the superconducting state for $0.05 < x < 0.25$. Meanwhile, long range FM order of Co $3d$-electrons develops above $x \approx 0.8$, with the transition temperature reaching 75 K in GdCoAsO (marked by FM$^\textup{{Co}}$). Moreover, for $0.7 \leq x < 1.0$, it appears that a magnetic reorientation transition occurs below $T^\textup{Gd}_\textup{N}$, as noted by $T^\ast$ in Fig.8.

Complex magnetic and superconducting phase diagrams have also been derived in other rare-earth iron pnictides. Resembling that of GdFe$_{1-x}$Co$_x$AsO, we found that, in CeFe$_{1-x}$Co$_x$AsO, the SDW transition of Fe $3d$-electrons is rapidly suppressed by Co substitution and superconductivity appears in a narrow doping range with a maximum $T_\textup{sc}$ of 13.5 K.~\cite{17tian} The Ce $4f$-electrons form a long-range AFM order at low temperatures, showing strong interactions with the $3d$-electrons on either the Fe- or the Co-rich side. In particular, a strong polarization effect induced by the ferromagnetism of Co $3d$-electrons on the magnetic moments of Ce is observed on the Co-rich side. Such a difference might result from the distinct magnetic coupling and electron hybridizations between the Gd- and Ce-compounds. In CeFe$_{1-x}$Co$_x$AsO, Ce atoms possess a small magnetic moment which can be modulated by the internal magnetic field of Co. However, the large moment of Gd in GdFe$_{1-x}$Co$_x$AsO is robust against the FM order of Co. The coupling between the Gd and Co lattices leads to ferrimagnetic behavior and also possible magnetic reorientation of Co at low temperatures. Since the FM order of Co is generally observed in the $Ln$CoAsO families,~\cite{27ohta2009magnetic} it is desirable to extend such measurements to other series, to study the interplay of $4f$- and $3d$-electrons and its influence on superconductivity and magnetism in iron pnictides.
\\
\section{CONCLUSION}
In summary, we have systematically studied the transport, magnetic and thermodynamic properties of a series of GdFe$_{1-x}$Co$_x$AsO ($0 \leq x \leq 1$) polycrystalline samples and presented a complete doping-temperature phase diagram. It is found that the SDW transition associated with the Fe $3d$-electrons is quickly suppressed upon Co substitution and superconductivity appears in a narrow doping range. The Gd is antiferromagnetically ordered over the whole doping range with a nearly unchanged N\'{e}el temperature. Above $x \approx 0.80$, the Co $3d$-electrons undergo a FM transition, whose transition temperature increases with increasing Co content and reaches 75 K in GdCoAsO. The magnetic susceptibility of Co-rich samples ($x > 0.8$) shows typical ferrimagnetic behavior, and can be well described in terms of the mean field model, while considering the AFM coupling between the Gd-species and the Co-species. These results suggest that the coupling between $3d$- and $4f$-electrons is enhanced on the Co-side in GdFe$_{1-x}$Co$_x$AsO, and such a strong coupling seems to disfavor superconductivity.

\begin{acknowledgments}
We would like to thank C. Geibel, E. D. Bauer, R. E. Baumbach, F. C. Zhang, Z.A. Xu, G. H. Cao and J. H. Dai for useful discussions. Work at Zhejiang University is supported by the National Basic Research Program of China (Nos.2009CB929104 and 2011CBA00103),the NSFC (Nos.10934005 and 11174245), Zhejiang Provincial Natural Science Foundation of China and the Fundamental Research Funds for the Central Universities. Work at Los Alamos National Lab was performed under the auspices of the US DOE and supported in part by the Los Alamos LDRD program. Work at The University of Texas at Dallas is supported by the AFOSR grant (No.FA9550-09-1-0384) on search for novel superconductors.
\end{acknowledgments}

\end{document}